# Geometrically-Controlled Microscale Patterning and Epitaxial Lateral Overgrowth of Nitrogen-Polar GaN

**Pietro Pampili[1], Vitaly Z. Zubialevich[1] and Peter J. Parbrook[1,2]**
[1]*Tyndall National Institute, University College Cork, Lee Maltings, Dyke Parade, Cork, Ireland*
[2]*School of Engineering, University College Cork, Western Road, Cork, Ireland*

**Abstract:**
In this study we report on a novel two-step epitaxial growth technique that enables a significant improvement of the crystal quality of nitrogen-polar GaN. The starting material is grown on 4° vicinal sapphire substrates by metal organic vapour phase epitaxy, with an initial high-temperature sapphire nitridation to control polarity. The material is then converted into a regular array of hexagonal pyramids by wet-etch in a KOH solution, and subsequently regrown to coalesce the pyramids back into a smooth layer of improved crystal quality. The key points that enable this technique are the control of the array geometry, obtained by exploiting the anisotropic behaviour of the wet-etch step, and the use of regrowth conditions that preserve the orientation of the pyramids' sidewalls. In contrast, growth conditions that cause an excessive expansion of the residual $(000\bar{1})$ facets on the pyramids' tops cause the onset of a very rough surface morphology upon full coalescence. An X-ray diffraction study confirms the reduction of the threading dislocation density as the regrowth step develops. The analysis of the relative position of the $000\bar{2}$ GaN peak with respect to the $0006$ sapphire peak reveals a macroscopic tilt of the pyramids, probably induced by the large off-axis substrate orientation. This tilt correlates very well with an anomalous broadening of the $000\bar{2}$ diffraction peaks at the beginning of the regrowth step.

**Main text:**
The growth of N-polar GaN has been intensively studied over the last two decades, especially for application in the field of high-frequency electronics, due to the disruptive approach in transistor scaling that N-polar III-nitrides enable [1]. It is now well understood that N-polarity can be controlled either by using the appropriate face of a polar substrate, such as the carbon face of SiC, or by applying a pre-growth high-temperature nitridation step to a sapphire substrate [2]. Early N-polar samples showed very rough surfaces, which was understood as due to reduced adatom mobility on N-polar surfaces and consequent random nucleation of hexagonal islands. An important breakthrough was the demonstration of smooth surface morphologies, which was made possible by the use of vicinal substrates with offcut angles as large as 4° [3, 4].

However, despite all these achievements, the crystal quality of N-polar GaN is still somewhat poorer than that of standard Ga-polar epilayers, with X-Ray rocking curve Full Widths at Half Maxima (FWHM) that are around 400 and 600 arcsec for symmetric (e.g. $000\bar{2}$) and skew-symmetric (e.g. $10\bar{1}\bar{2}$ and $10\bar{1}\bar{1}$) reflections, respectively, for most of the best samples reported in literature (see for example [4] for sapphire and [5] for SiC substrates; more references can be found also in [1]).

In the past, several growth techniques have been developed to reduce the threading dislocation density of heteroepitaxial films. An effective approach is based on the initial formation of microscale structures having slanted sidewalls, followed by the subsequent coalescence of these structures into a smooth, higher-quality epilayer. During the coalescence phase, the dislocations that cross any of the slated facets are forced to bend outwards, which increases the likelihood of meeting and annihilating with other nearby dislocations. Examples of this approach in III-nitrides are the Facet Initiated Epitaxial Lateral Overgrowth (FIELO) technique, whereby slanted sidewalls occur spontaneously with Hydride Vapour Phase Epitaxy (HVPE) [6], and the Facet Controlled Epitaxial Lateral Overgrowth (FACELO) technique, whereby specific growth conditions are needed to induce slanted sidewalls in Metal Organic Vapour Phase Epitaxy (MOVPE) [7]. We also developed a similar approach with lateral expansion of nanocolumns [8]. Despite the evident success of this approach for Ga-polar materials, there is currently no report of similar techniques for N-polar materials, which is the topic of this study.

A striking difference between Ga- and N-polar GaN is that, while the former is chemically very stable, the latter can be easily wet-etched in KOH or other hydroxide-containing solutions, which makes this the simplest and most widely used way to discriminate between the two polarities [9]. It is also well-known that during etching the $-c$ face quickly converts into a random distribution of hexagonal pyramids with slanted facets oriented along $\{1\bar{1}0\bar{1}\}$ planes, a fact that has been used, for example, to roughen the bottom N-polar face of laser-lifted-off LEDs to increase light-extraction efficiency [10]. These naturally occurring structures would be the perfect candidates for a FACELO-like technique for N-polar materials, with the additional advantage of not requiring any particular optimization of the growth conditions to force their appearance. Wet-etching is also a very simple and low-cost procedure that does not induce crystal damage as opposed to plasma-based dry-etching. Unfortunately, the fact that the pyramids form with a large distribution of heights makes it virtually impossible to convert them back into a smooth epilayer during the coalescence phase. Hence, the development of a new strategy for the control of the pyramid geometry during etching is crucial.

From the observation that the wet-etch of N-polar materials continues until the epilayer is completely dissolved, and does not stop even after there are no more $-c$ facets left and the surface is fully covered in pyramids, it is often concluded that both $(000\bar{1})$ and $\{1\bar{1}0\bar{1}\}$ planes are attacked, although possibly at different rates [10]. However, a more accurate analysis of the etch mechanism reveals that the etching in the direction perpendicular to the $\{1\bar{1}0\bar{1}\}$ planes is only apparent.

In reality, we have found that, in a wide range of temperatures and concentrations, the $\{1\bar{1}0\bar{1}\}$ planes are very stable in KOH solutions, and the hexagonal pyramids, once formed, can only be attacked from their tops, where a nanoscale $-c$ facet is continuously recreated and etched away, as schematically shown in Figure 1. Consequently, because the $\{1\bar{1}0\bar{1}\}$ planes can only be etched 'laterally' from the $[000\bar{1}]$ direction, the deposition of a protective cap is sufficient to stabilize them.

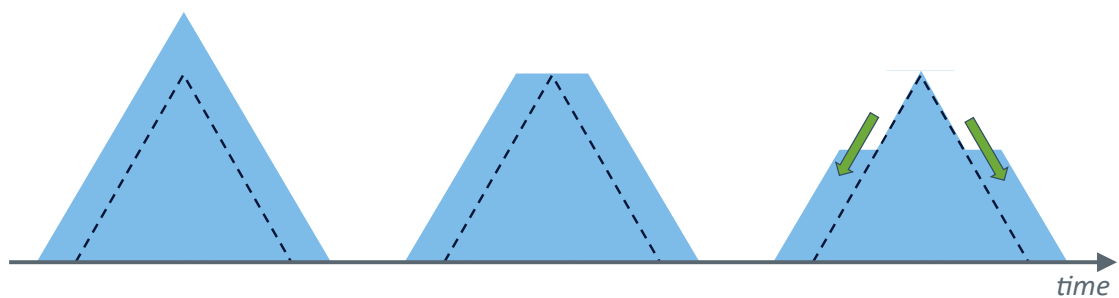

*Figure 1. Etching mechanism of N-polar GaN pyramids in KOH solutions. The $\{1\bar{1}0\bar{1}\}$ oriented sidewalls are not etched perpendicularly, but laterally attacked from the top, in an atomic layer-by-layer process. As $(000\bar{1})$ oriented nanoscale facets of just a few atoms are continuously recreated by the etchant solution, the process continues indefinitely.*

This is clearly demonstrated in the Scanning Electron Microscopy (SEM) images shown in Figure 2, where a N-polar GaN sample, first patterned with SiN caps and then exposed to a KOH-containing solution, is shown at two different etch times. While the area in-between the caps did initially convert into a large number of small pyramids of random heights (Figure 2a), the material below the caps was preserved, and eventually evolved into a regular array of larger pyramids. Moreover, after meeting with neighbouring pyramids, their slanted sidewalls stopped each other's progression and formed a very stable final configuration (Figure 2b). Hence, this self-limiting behaviour allows also for etch-depth control by the choice of appropriate cap diameters and pitch-distances. The so-obtained structure is the ideal starting template for a FACELO-like regrowth technique to improve crystal quality of N-polar GaN.

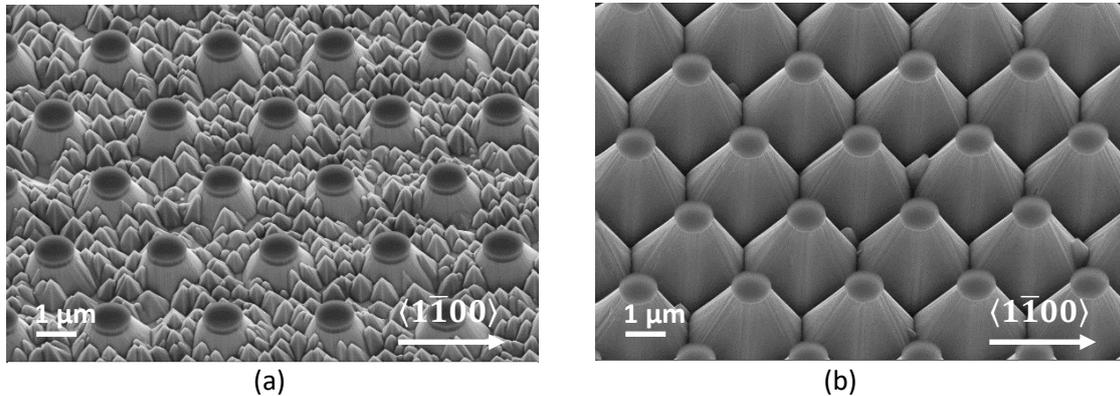

(a)  (b)

Figure 2. SEM image (45° tilted) of a N-polar GaN sample patterned with SiN caps and etched in AZ 400K (a) after 15 minutes at room temperature, and (b) after 3 hours at 80°C. While the resulting pyramids are mostly $\{1\bar{1}0\bar{1}\}$ oriented, some deviations from crystallographic planes are induced by the circular caps, particularly at the sidewalls' edges.

In preparation for the regrowth experiments, we fabricated a set of nominally identical templates, similar to the one shown in Figure 2b. The starting planar material was grown in a vertical 3x2" Close Coupled Showerhead Aixtron MOVPE reactor using the same approach reported in [11], in which a vicinal sapphire substrate, with offcut angle of 4° towards the sapphire *a*-direction, is pre-conditioned in a high-temperature nitridation step, and the subsequent growth is carried out in a standard 2-temperature sequence, although with the nucleation done at a higher temperature compared to typical Ga-polar growth. In total, about 2.2 µm of material was grown. Subsequently, ~100 nm of SiN was blanket-deposited on the wafers using Plasma-Enhanced Chemical Vapour Deposition (PECVD), and, on top of it, a regular array of Pd discs was created by metal evaporation and an optical-photolithography lift-off step. The Pd discs were then used as a hard mask for a dry-etch step to remove all unprotected SiN. Finally, the Pd was dissolved in aqua regia to expose the newly created SiN caps, which had a nominal diameter of 1 µm, and were arranged in a hexagonal pattern with a 3-µm pitch distance. Critically, the caps were oriented with respect to the epilayer crystal lattice so that neighbouring caps were aligned along $\langle 1\bar{1}00 \rangle$ directions. The samples were then wet-etched for 3 hours in the developer AZ 400K, which is a buffered KOH solution, at a temperature of 80°C to form the patterned templates.

To study how coalescence evolves under different growth conditions, we conducted three series of regrowth experiments. Each series consisted of subsequent runs done at the same nominal conditions, but with increasingly longer growth times. In a previous study on Ga-polar GaN nanocolumns [12], we found that low-temperature growth in nitrogen favours nanocolumn expansion, while more standard growth in hydrogen and at higher temperatures leads to infilling from the bottom. Based on this observation, we carried out a similar comparison for N-polar growth.

In the first series, samples were grown under nitrogen, at a temperature of 950°C (determined using a calibrated pyrometer), chamber pressure of 100 mbar, and a V/III ratio of 1,000. As expected, this resulted in a clear expansion of the pyramids, and no noticeable infilling of the areas in between them. However, as shown in Figure 3, the expansion was not uniform along the pyramids' height.

During the early stages of regrowth, only the tops were significantly affected, while the bottom parts showed minor reshaping. As expansion continued, more and more sections towards the bottom of the pyramids started to follow, which eventually resulted, shortly before full coalescence, in the formation of rather steep sidewalls approaching *m*-plane orientation. An interesting effect of this type of growth evolution is that the pyramid tops, initially fully covered by the SiN caps, expanded outwards forming quite large and crystallographically smooth $-c$ facets. It is worth highlighting that growth on top of these now perfectly on-axis $(000\bar{1})$ tops would result in the typical rough morphology observed in non-offcut samples. The fact that they remained smooth for most of the regrowth proves that only lateral expansion was happening at these stages. However, as the growth progressed further, and the lateral expansion became geometrically constrained, the surface supersaturation on the tops started to increase, which initially led to occasional nucleation and formation of sparse hillocks (not shown), and upon full coalescence, to the abrupt onset of the rough surface morphology shown in Figure 3d.

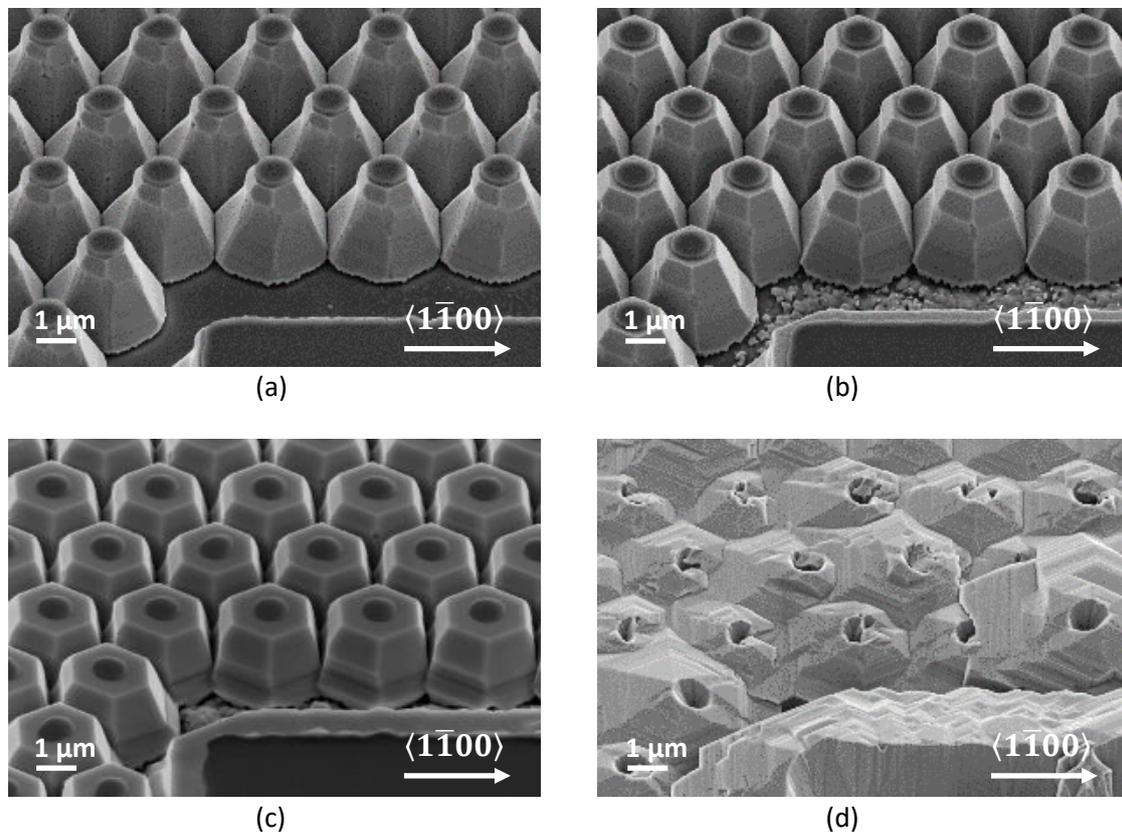

*Figure 3. Series of overgrowth experiments conducted in nitrogen, at 950°C. Nominal overgrowth thickness on planar reference: (a) 250 nm, (b) 750 nm, (c) 1,750 nm and (d) 2,750 nm.*

In the second series of experiments, the samples were grown in a hydrogen ambient, at a temperature of 1040°C, pressure of 150 mbar, and V/III ratio of 1,000. Compared with growth under nitrogen, the expansion of the $(000\bar{1})$ tops was much less pronounced, and the areas at the base of the pyramids started infilling from the beginning, as shown in Figure 4. The sidewalls initially expanded maintaining mostly $\{1\bar{1}0\bar{1}\}$ orientation, but some flatter facets (most likely $\{1\bar{1}0\bar{2}\}$) eventually formed around the tops (Figure 4b), and became dominant just before coalescence (Figure 4c). This indicates that $\{1\bar{1}0\bar{1}\}$ planes grow faster than $\{1\bar{1}0\bar{2}\}$ planes, and these faster than the $(000\bar{1})$ plane, which is the opposite order of what is observed in-metal polar III-nitrides [13].

This different type of growth evolution made it possible a gradual coalescence of the pyramids, without the sudden increase of supersaturation on the $(000\bar{1})$ tops that had led to rough surface morphology on the previous experiment. Most of the areas in-between the caps became rather smooth, but, as the growth progressed even further above the original height of the pyramids, the material that started to grow on top of the caps became rough. In some cases, we even observed Ga-polar nucleation, instead of expansion and overgrowth of N-polar material. This was confirmed by wet-etch in AZ 400K, which removed all the overgrown N-polar GaN, and restored the original geometry of the as-wet-etched templates, but could not dissolve the material grown on top of some of the caps.

However, regardless of the presence of any SiN cap, both series of experiments clearly showed that no growth on top of the pyramids is possible, unless the supersaturation is increased above a certain level, which in the conditions explored here, happens only when all faster growing planes are almost completely consumed.

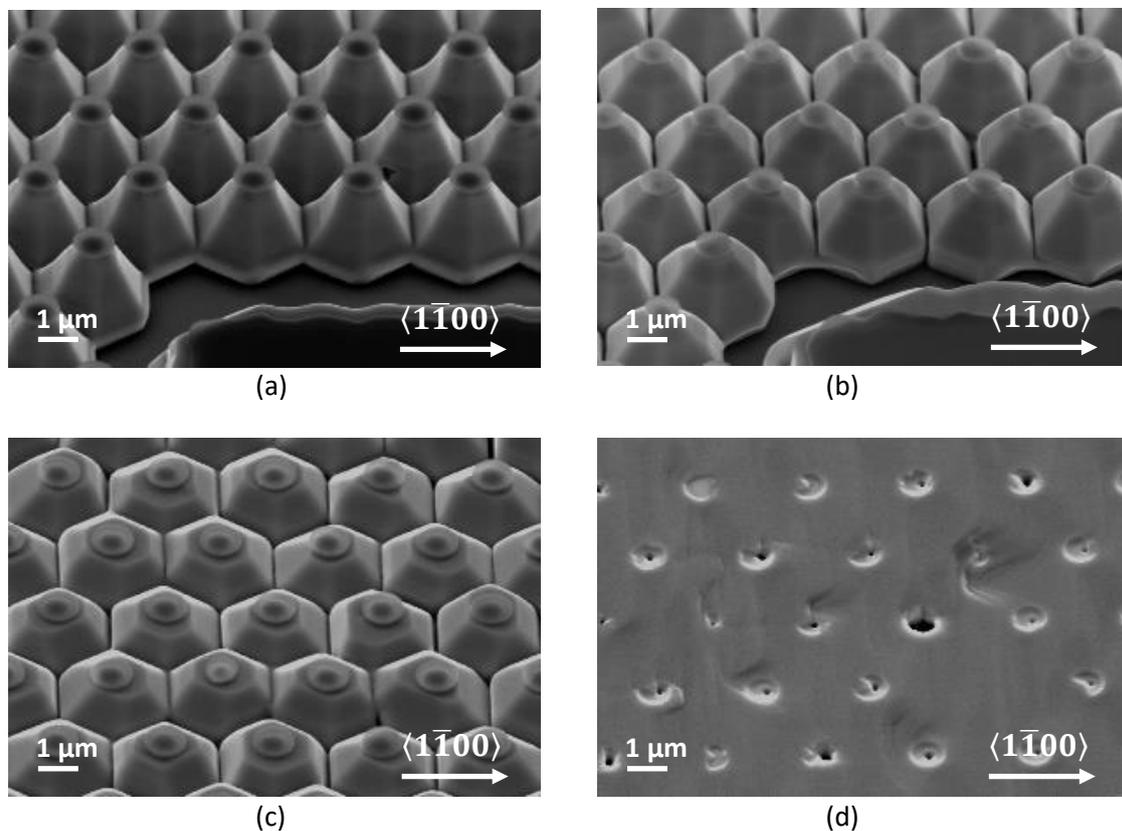

*Figure 4. Series of overgrowth experiments conducted in hydrogen, at 1,040°C. Nominal overgrowth thickness on planar reference: (a) 250 nm, (b) 500 nm, (c) 1,000 nm and (d) 2,000 nm.*

For this reason, a third series of regrowth runs in hydrogen was conducted at the same nominal conditions as before, but on templates whose pyramids had previously been removed of their SiN caps by wet-etch in buffered HF solution. In addition to that, we increased the number of runs to study the growth evolution in more detail. As can be seen in Figure 5, this time the coalescence phase concluded with the formation of a smooth surface of comparable morphology to that of the original as-grown epilayer. Even at a larger scale, when observed under Nomarski phase contrast, the surface appeared smooth and free from hexagonal islands or other macroscopic defects, as shown in Figure 5h. Immediately after coalescence, the surface morphology was temporarily dominated by parallel striations oriented perpendicularly to the offcut direction, and whose average distance was exactly half of the pyramid pitch-distance. From this we can conclude that these striations are formed when the slowly expanding pyramid's tops finally meet. Being crystallographically $(000\bar{1})$ oriented, the tops are necessarily tilted by 4° with respect to the sample face, which was grown on a vicinal substrate off-axis by the same angle. When neighbouring tops meet, they have hence slightly different heights, which creates a sort of macroscopic step bunching. However, without any disturbance from the SiN caps, the step edges were free to move as step-flow growth was re-established, and eventually redistributed uniformly, restoring a 4° off-axis surface, smooth and free from any major striations.

In order to gain insight into the evolution of the threading dislocation density during the coalescence phase, the samples of the third series were also analysed with X-Ray Diffraction (XRD) using a Malvern Panalytical X'Pert diffractometer. Peak broadening in symmetric (skew-symmetric) reflections is commonly used as a proxy for the density of threading dislocations with a screw (edge) component, respectively [14]. For this reason, FWHMs of $000\bar{2}$ and $10\bar{1}\bar{1}$ reflections were measured using a PIXcel detector in open-detector mode. At an initial analysis, some FWHM variations were observed at different, but nominally equivalent, azimuthal angles $\varphi$, probably induced by the presence of the large offcut angle at a particular $\varphi$. For this reason, the offcut direction, which for all the samples here studied corresponded to a sapphire $a$-direction—and hence to a GaN $m$-direction—was defined as $\varphi = 0°$. Then, for consistency, all $000\bar{2}$ reflections were measured at $\varphi = 90°$, and all $10\bar{1}\bar{1}$ skew-symmetric reflections at $\varphi = 30°$.

As shown in Figure 6a, the as-wet-etched starting template had FWHMs of 529 and 552 arcsec, for $000\bar{2}$ and $10\bar{1}\bar{1}$, respectively, which are similar to the values commonly reported in literature. As the coalescence phase progressed, the $10\bar{1}\bar{1}$ FWHMs decreased monotonically, demonstrating a clear edge-dislocation reduction. After regrowth of nominally 3 µm of material, the FWHM decreased to 329 arcsec. For the $000\bar{2}$ reflections, a different type of evolution was observed, whereby the FWHMs initially increased over 650 arcsec, and only after 1 µm of nominal growth, started to decrease and to follow a trend similar to the one observed for $10\bar{1}\bar{1}$, which demonstrate a comparable reduction also for screw dislocations. At the end of the experiment, the measured $000\bar{2}$ FWHM was 321 arcsec. It is worth noting that the final FWHM values here reported represent a rather large underestimation of the crystal quality of the coalesced epilayer, as both the newly grown and the underlying original material are probed during these XRD measurements. Nevertheless, the steady FWHM reduction after 1 µm of nominal growth is a clear signature of threading dislocation density reduction in N-polar epitaxial lateral overgrowth. Even though the geometry of the starting patterned template was not optimized, the resulting FWHM values are already significantly smaller than those typically reported in literature for N-polar GaN on foreign substrates.

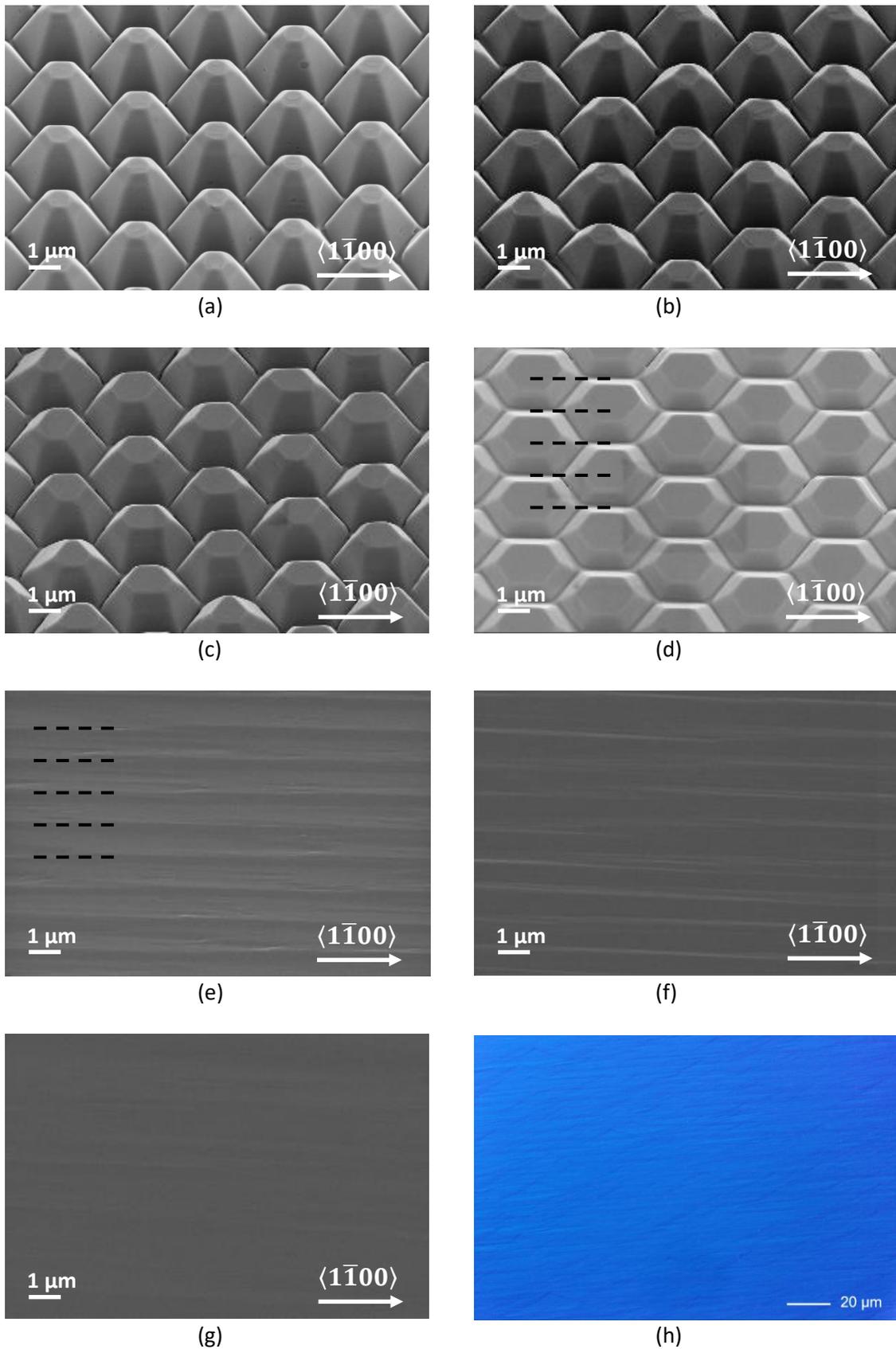

*Figure 5. Series of overgrowth experiments conducted in nitrogen, at 950°C, on wet-etched templated with removed SiN caps. Nominal overgrowth thickness on planar reference: (a) 250 nm, (b) 500 nm, (c) 750 nm, (d) 1,000 nm, (e) 1,500 nm, (f) 2,000 nm and (g) 3,000 nm. (h) Nomarski microscopy image taken after overgrowth of 3,000 nm of material. The formation of the striations and their mutual distance is highlighted by the dashed lines in (d) and (e).*

In heteroepitaxial growth, lattice mismatch between substrate and epilayer induces strain; however, while in on-axis epitaxy only *a*-lattice parameter differences matter, for growth on vicinal substrates also *c*-lattice parameter mismatch plays a role. In fully strained epilayers this in known to cause the so-called Nagai tilt [15, 16]. To investigate the presence of any tilt in our samples, we collected a series of reciprocal space maps of the GaN-$000\bar{2}$ and sapphire-0006 nearby reflections, both at $\varphi = 0°$ and $\varphi = 90°$, as shown for example in Figure 6c and d for one of the samples. The epilayer relative tilt $\alpha_\varphi$ at the azimuthal angle $\varphi$, was calculated as the offset difference between the two peaks:

$$\alpha_\varphi = (\omega_{epi,\varphi} - \theta_{epi,\varphi}) - (\omega_{sub,\varphi} - \theta_{sub,\varphi}),$$

where the subscripts *epi* and *sub* refer to epilayer and substrate, respectively. The evolutions of both $\alpha_{90°}$ and $\alpha_{0°}$ tilts are shown in Figure 6b. As expected, there is almost no tilt along the direction $\varphi = 90°$ perpendicular to the offcut, but a tilt along $\varphi = 0°$ is clearly noticeable, although of much smaller intensity than Nagai theory would predict, which indicates almost complete epilayer relaxation. Interestingly, the evolution of the tilt in the offcut direction correlates very strongly with the anomalous increase of the $000\bar{2}$ FWHMs at the initial stage of the coalescence phase shown in Figure 6a. This seems to suggest that nucleation of new screw dislocations might indeed be happening at the very early stages of the coalescence, caused either by the epilayer tilt, or, possibly, by its pyramid-to-pyramid variations. However, further work is still necessary to better understand the interplay between macroscopic tilt induced by the growth on vicinal substrates, and microscopic tilt due to screw dislocations.

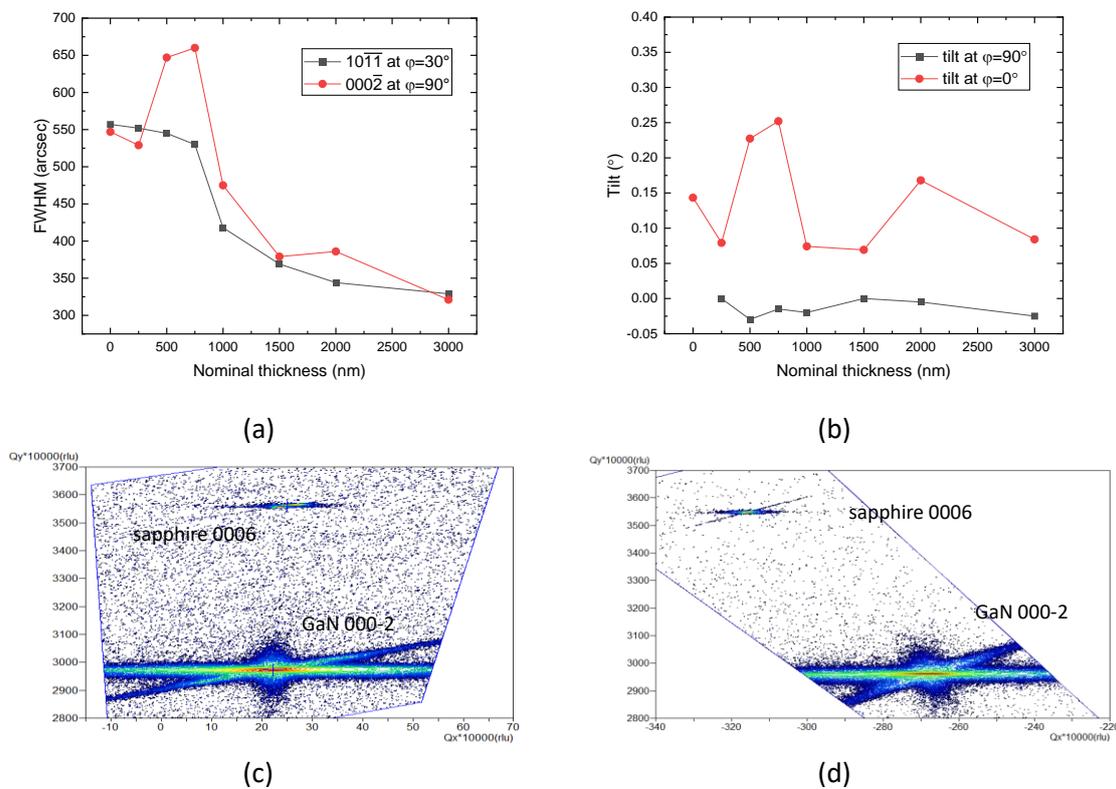

Figure 6. Evolution of the coalescence phase as a function of nominal overgrowth thickness for: (a) FWHMs of $000\bar{2}$ and $10\bar{1}\bar{1}$ reflections; and (b) epilayer tilt in the directions corresponding to $\varphi = 90°$ and $\varphi = 0°$. Example of two reciprocal space maps collected, for the same sample, around GaN-$000\bar{2}$ and sapphire-0006 reflections and: (c) $\varphi = 90°$, with offcut accommodated by $\chi$; and (d) $\varphi = 0°$, with offcut accommodated by $\omega$ offset.

In conclusion, we have demonstrated—for the first time in N-polar materials—a significant reduction of threading dislocations with epitaxial lateral overgrowth of a patterned $(000\bar{1})$ GaN layer. The starting templates were prepared by wet-etching in KOH-containing solution, and consisted in a regular array of microscale hexagonal pyramids with sidewalls approximately oriented along the $\{1\bar{1}0\bar{1}\}$ planes. The coalescence of so-created pyramids into a final layer of adequately smooth surface morphology required the use of growth conditions that preserved the sidewalls' orientation over expansion of the $(000\bar{1})$-oriented pyramid tops. Although further optimization of the process is still needed, this proof-of-concept demonstration shows that there is room for significant improvement of current N-polar GaN crystal quality, which would be of great benefit for the development of new and improved N-polar devices.

**Acknowledgements:**

This project has received funding from the European Union's Horizon 2020 research and innovation program under the Marie Skłodowska-Curie grant agreement No. 898704.
Part of the work was also supported by Science Foundation Ireland (SFI) under grant numbers 12/RC/2276_P2 and 21/FFP-A/9014. Open access funding was provided by IReL.

10. Wang L, Ma J, Liu Z, et al (2013) N-polar GaN etching and approaches to quasi-perfect micro-scale pyramid vertical light-emitting diodes array. J Appl Phys 114:133101. https://doi.org/10.1063/1.4823849

11. Pristovsek M, Furuhashi I, Pampili P (2023) Growth of N-Polar (000-1) GaN in Metal–Organic Vapour Phase Epitaxy on Sapphire. Crystals 13:. https://doi.org/10.3390/cryst13071072

12. Zubialevich VZ, Pampili P, Parbrook PJ (2020) Thermal Stability of Crystallographic Planes of GaN Nanocolumns and Their Overgrowth by Metal Organic Vapor Phase Epitaxy. Cryst Growth Des 20:3686–3700. https://doi.org/10.1021/acs.cgd.9b01656

13. Singh SM, Zubialevich VZ, Parbrook PJ (2024) Surface Morphology Evolution of AlGaN Microhoneycomb Structures during Epitaxial Overgrowth. Phys Status Solidi B Basic Res. https://doi.org/10.1002/pssb.202300471

14. Moram MA, Vickers ME (2009) X-ray diffraction of III-nitrides. Rep Prog Phys 72:. https://doi.org/10.1088/0034-4885/72/3/036502

15. Nagai H (2003) Structure of vapor-deposited $Ga_xIn_{1-x}As$ crystals. J Appl Phys 45:3789. https://doi.org/10.1063/1.1663861

16. Huang XR, Bai J, Dudley M, et al (2005) Epitaxial tilting of GaN grown on vicinal surfaces of sapphire. Appl Phys Lett 86:211916. https://doi.org/10.1063/1.1940123
10